# Coherent spin qubit transport in silicon

J. Yoneda[1]*, W. Huang[1], M. Feng[1], C. H. Yang[1], K. W. Chan[1], T. Tanttu[1], W. Gilbert[1], R. C. C. Leon[1], F. E. Hudson[1], K. M. Itoh[2], A. Morello[1], S. D. Bartlett[3], A. Laucht[1], A. Saraiva[1], A. S. Dzurak[1]*

[1]School of Electrical Engineering and Telecommunications, The University of New South Wales, Sydney, NSW 2052, Australia

[2]School of Fundamental Science and Technology, Keio University, Yokohama, Japan

[3]Centre for Engineered Quantum Systems, School of Physics, University of Sydney, Sydney, NSW 2006, Australia

*Correspondence to: j.yoneda@unsw.edu.au, a.dzurak@unsw.edu.au

**A fault-tolerant quantum processor may be configured using stationary qubits interacting only with their nearest neighbours, but at the cost of significant overheads in physical qubits per logical qubit. Such overheads could be reduced by coherently transporting qubits across the chip, allowing connectivity beyond immediate neighbours. Here we demonstrate high-fidelity coherent transport of an electron spin qubit between quantum dots in isotopically-enriched silicon. We observe qubit precession in the inter-site tunnelling regime and assess the impact of qubit transport using Ramsey interferometry and quantum state tomography techniques. We report a polarization transfer fidelity of 99.97% and an average coherent transfer fidelity of 99.4%. Our results provide key elements for high-fidelity, on-chip quantum information distribution, as long envisaged, reinforcing the scaling prospects of silicon-based spin qubits.**

Harnessing the full potential of quantum computers requires the use of quantum error correction [1,2]. A popular strategy based on the 2D surface code [3] requires only nearest-neighbour operations between physical qubits with a very lenient error threshold. These advantages, however, come at the cost of severe overheads in the number of physical qubits per logical qubit and the need for resource-intensive magic state distillation to achieve universal quantum logic. Long-range interactions offer the potential to substantially reduce these overheads, with several recent innovations in quantum architectures [4–7] exploiting long-range operations to perform error correction with fixed overheads, as well as fault-tolerant logic without magic state distillation. Non-local quantum operations can also provide advantages for near-term, non-error-corrected systems [8].

Furthermore, in semiconductor quantum processors, where the physical qubits have nanometre-scale footprint, non-local operations can help reduce the density of control lines [9], or allow interspersing of classical electronics between densely-packed qubit modules [10,11]. Although the demonstrated performance of prototypes based on silicon quantum dot qubits [12–18] suggests this system could be scaled up by leveraging industrial semiconductor technology [19–21], serious challenges still lie ahead for a dense array of stationary qubits with individualised control circuitry [22]. Benefits of incorporating qubit transport in the architecture have therefore been widely recognized [9–11,23,24]. Strategies for quantum information transfer in semiconductor spin qubits include sequential application of spin SWAP gates [25–27], coherent coupling of stationary qubits mediated by flying qubits such as photons in a cavity [28–30] or, as proposed in the literature [31,32] and explored here experimentally, physically transporting the particle that harbors the quantum information from one site to another [33–37]. However, the impact and error caused by the qubit transport process, which has so far been discussed theoretically [38–42], needs to be elucidated before real progress can be made on mobile qubit architectures.

In this work, we investigate how a single electron spin can be coherently transported within a silicon quantum-dot system. We develop a Ramsey spectroscopy-like technique that allows for a full characterization of the spin-qubit dispersion in tunnel-coupled quantum dots. We test the phase coherence of the transport process by performing quantum state tomography on a post-transfer electron spin. By transferring an electron repeatedly between two sites we obtain a spin polarization fidelity of $(99.9703 \pm 0.0007)$% (average of ↑ and ↓) and an average gate fidelity of the coherent transfer process of $(99.36 \pm 0.05)$%. By measuring the spin coherence after multiple transfer cycles with the phase evolution time fixed, we distinguish the impact of the phase error per transfer event from the usual temporal dephasing. This transfer method can be extended to longer quantum-dot chains by sequencing it from one site to the next in a bucket-brigade manner, offering micron-scale on-chip quantum links for silicon spin-qubit architectures.

We host our spin qubit in a pair of metal-oxide-semiconductor (MOS) quantum dots [13] in isotopically-enriched silicon (Fig. 1a). We can move the single electron between sites A and B by biasing the voltages applied to the surface gate electrodes (Fig. 1b). In what follows, we sweep the gate voltages along a detuning axis $\varepsilon$ (see Fig. 1c), with its value given by the gate B voltage with respect to the interdot transition (which is precisely determined experimentally below). This changes the energy difference between the states localized in individual sites. We can sense the charge configuration (Fig. 1c) by using a single-electron transistor (SET) as an electrometer. We initialize and read out the spin state based on spin-selective tunnelling from site A to the reservoir, in combination with charge sensing (Supplementary Information).

We can manipulate the qubit via electron spin resonance (ESR) with the frequency of the control a.c. magnetic field tuned to the electron-spin Zeeman splitting. The Zeeman energy is often site dependent in silicon MOS quantum dots due to the interplay between the spin-orbit interaction and the confinement electric field [13,17,19]. Consistent with this, we measure a clear resonance frequency shift of roughly 30 MHz when $\varepsilon$ is changed (Fig. 1d). The observed smoothness of the transition between the two frequencies indicates that the electron wavefunctions are strongly hybridized between sites by the interdot tunnel coupling and the so-called bonding state is formed.

We first confirm that the polarization of the spin can be transported between sites with high fidelity. The main concern would be that the energy levels of opposite spins in sites A and B would eventually match when the interdot detuning ε becomes equal to the Zeeman splitting, facilitating a spin-flip tunnelling process from site A to B due to the spin-orbit field generated by the electron movement or a small site difference in spin quantization axes [13,43,44]. We avoid the formation of these degeneracy points by enhancing the tunnel coupling above the Zeeman energy (~28 GHz). A large tunnel coupling will also suppress state leakage due to non-adiabatic tunnelling [41]. To amplify the polarization error to a measurable level, we repeatedly transfer the spin (initialized in either ↓ or ↑) between the sites (Fig. 1e). The detuning ramps are applied at 56 ns intervals to ensure the spin is transferred to the other site (Fig. S1). From the analysis (Supplementary Information), we obtain the polarization transfer fidelities of $(99.9514^{+0.0008}_{-0.0017})\%$ and $(99.9892^{+0.0008}_{-0.0008})\%$ for the ↑ and ↓ cases, respectively (Fig. 1f). Here the error bars denote a 1σ confidence interval from the fit. The spin polarization fidelity is high enough that spin flips do not play a role in the following experiments.

We now address whether the coherence is retained when the qubit is moved across sites by employing a Ramsey-type protocol (Fig. 2a). We first prepare a spin in an equal superposition of ↑ and ↓ states using a π/2 ESR pulse (on resonance with the Larmor frequency at site A). We then pulse the detuning ε from $\varepsilon_1$ (in site A) to $\varepsilon_2$ (either in site A or B), for a duration of $t_{\text{dwell}}$, on a nanosecond timescale. The phase acquired during the round trip to $\varepsilon_2$ is then projected to spin polarization by a second π/2 ESR pulse in site A. Figure 2b plots the final ↑ probability ($P_\uparrow$) after this coherent tunnelling spectroscopy. The oscillation of the probability $P_\uparrow$ as a function of time $t_{\text{dwell}}$ spent at detuning $\varepsilon_2$ is visible, irrespective of how deeply we pulse $\varepsilon_2$, suggesting the whole process is phase coherent. Importantly, the fringe frequency starts to rapidly change for $\varepsilon_2 > 0$ and saturates at around 30 MHz (consistent with the qubit resonance frequency difference between sites), indicating that the electron is indeed completely transferred to site B in the saturated region ($\varepsilon_2 > 5$ mV). This demonstrates that the spin can be shuttled to a different site and back while maintaining phase coherence.

The coherent tunnelling spectroscopy technique described above accurately measures the qubit precession frequency as a function of the gate voltage (see Fig. 2c and Fig. S2, where the analysed ESR frequency is also plotted for comparison). This allows us to establish a detailed understanding of the qubit dispersion in our tunnel-coupled quantum-dot system. The qubit frequency dependence $f_Q(\varepsilon)$ can be fitted to a simple model with a single orbital per dot (disregarding orbital and valley excitations, see Supplementary Information), which determines the tunnel coupling to be 103.8 ± 1.5 GHz and the interdot transition to occur at a gate B voltage of 968.85 ± 0.04 mV, defining what is experimentally considered to be the point where $\varepsilon = 0$ mV. This model also allows us to precisely calculate the wavefunction hybridization for a given gate voltage condition (Fig. 2c, right ordinate). We note that the qubit frequency is best fit with a small spin-dependence in the interdot tunnel coupling due to spin-orbit interaction (Supplementary Information). Furthermore, we discover a detuning spot (roughly around $\varepsilon$ = -7 mV) where the qubit frequency is first-order insensitive to detuning fluctuations due to charge noise, as a result of competition between the Stark shift and the tunnelling hybridization.

It is worth noting that we can complete the qubit shuttling within nanoseconds, several orders of magnitude faster than the qubit dephasing time. To illustrate the potential use of this as part of qubit control protocols, we demonstrate a gate-voltage-controlled phase gate in Fig. 2d, where the phase is acquired mostly due to the site-dependence of qubit frequency and the hybridization effect due to interdot tunnelling rather than the intradot Stark shift [11,12]. We find that the phase accumulates at ~30 MHz consistently down to an 8 ns dwell time (limited by our control hardware). Similarly, we can use the frequency difference between sites to tune the qubit in and out of resonance with regard to a fixed ESR control tone (see Fig. S3), useful for qubit addressing in an always-on microwave control field [45].

We further assess the influence of the tunnelling process on the qubit by performing quantum state tomography for the spin state with and without a site-to-site transfer. As schematically shown in Fig. 3a, we first prepare a $|+y\rangle$ state in site A ($\varepsilon = -10$ mV) using a π/2 ESR pulse. We then either transfer the electron to site B ($\varepsilon = +10$ mV) or leave it idling in site A for the same amount of time as the transfer would take. We finally measure the state along ten different axes (see Supplementary Information for details) and reconstruct the spin density matrix (Fig. 3b) using the maximum likelihood estimation technique [13–15]. The state after a transfer is well approximated by a pure, equal superposition of ↑ and ↓ states (i.e. a Bloch vector on the Bloch sphere's equator). This further verifies that the site-to-site qubit transfer process can be well-approximated by a unitary phase rotation gate. Its rotation angle $\Delta\varphi$ can be related to the $\varepsilon$-dependent qubit frequency (dominated by the site-dependent Zeeman energy); note that we describe the qubit at all times in the rotating frame of the driving microwave, and that $\Delta\varphi$ is affected by the synchronization between the electron tunnelling time and the instant at which we switch between the resonance frequencies in sites A and B. Comparing the reconstructed spin state after a transfer with the idealized case – a pure state obtained after applying an ideal phase gate to an exact $|+y\rangle$ state – we estimate a state fidelity of $(98.7^{+0.6}_{-0.8})\%$ in the absence of errors in the state preparation and measurement (SPAM). Alternatively, the spin state without a transfer has a fidelity of $(97.5^{+0.5}_{-0.8})\%$ after correcting for SPAM errors. The results indicate that the transfer process is highly coherent and that any difference between the two states is below the sensitivity of this measurement.

In order to quantify the small phase error of the transport process in the presence of SPAM errors, we employ a sequence where the transport ramp pulses are repeated many times between state preparation and measurement, and evaluate the remaining spin coherence using a Ramsey-interference technique. This protocol amplifies errors, leading to a decay of the phase oscillation amplitude with the number of transfer cycles, $n$. If the error probability of consecutive transfers is uncorrelated, the amplitude decay will be exponential. We first investigate the case of round trips (Fig. 4a). The qubit is prepared in site A, transferred back and forth an even number ($n$) of times between sites A and B ($\varepsilon$ = -10 mV and 10 mV) before it is measured in the original site, A. To change the projection axis, the spin state is rotated around various in-plane axes by changing the microwave phase $\phi$ of the second $\pi/2$ pulse. The fringe amplitude as a function of $\phi$ (Fig. 4b) reflects the spin phase coherence after the ramps and decreases when the number of transfers $n$ is increased as well as the phase evolution time $T_{evol}$ (the interval between the preparation and projection ESR pulses, see Fig. 4a). From the exponential decay rate of the fringe amplitude as a function of $n$ (Fig. 4c), we extract the coherence loss per transfer, $p$, of $(2.10^{+0.13}_{-0.09})\%$. We can extend this scheme to the odd-$n$ transfer case, in which the qubit phase is projected while in site B (with a microwave tone tuned for site B). Despite a slightly increased pulse complexity, the obtained value of $p$ is almost identical, $(2.01^{+0.24}_{-0.20})\%$ – see Fig. S4.

The coherence loss extracted above is a combination of the temporal dephasing of a freely precessing spin and the errors introduced by the transfer process. This is because the phase evolution time $T_{evol}$ increases by 56 ns per transfer in the above protocol and the temporal dephasing rate for a $T_2^*$ = 20 µs could be 0.3%. Instead, we can estimate the error induced by the transfer process *only* by using a slightly modified sequence in which $T_{evol}$ is fixed as $n$ is increased. The fringe decay rate (orange data in Fig. 4c) yields a coherence loss due to the transfer process, $p = (1.80^{+0.17}_{-0.16})\%$. A similar value of $p = (1.88^{+0.17}_{-0.16})\%$ is obtained for data where a different fixed value of the phase evolution time is used, see red data in Fig. 4c. The poorer of the two corresponds to a transfer fidelity of $(99.36 \pm 0.05)\%$, expressed in terms of an average gate fidelity of the transfer process (see Supplementary Information).

We attempt to improve the transfer fidelity, as well as investigate the noise spectrum, by including a refocusing pulse in our ramp sequence [46]. We adopt the protocol shown in Fig. 4d, where a decoupling $\pi$ pulse is applied between two identical series of transfer ramps. The echo fringes (Fig. 4e) are measured by sweeping the angle $\phi$ of the projection axis, revealing that the fringe phase does not change with the transfer cycles, as expected. The amplitude decay of the echo fringes as a function of $n$ (Fig. 4f) yields $p = (1.41^{+0.12}_{-0.06})\%$. While this provides some improvement in coherence, it means that the dominant part of phase error induced in the transfer process is not refocused.

This inefficiency of the dynamical decoupling pulse suggests that the underlying mechanism for the transfer-induced coherence loss $p$ ~2% is not dominated by either the slow spontaneous flips of the residual $^{29}$Si nuclear spins [37] or conventional charge-noise-induced dephasing with a $1/f$-type spectrum e.g. fluctuation in quantum dot levels [12] – see also Fig. S5. The transfer ramp used is slow enough to guarantee that no orbital or valley excitations may occur [41,42] – indeed, studying the influence of the detuning ramp rate, we verify that a slower ramp only degrades the transfer fidelity (Fig. S6). The linear increase in $p$ as a function of ramp time (~1/2.0 $\mu s^{-1}$) may be semi-quantitatively explained by enhanced dephasing and/or diabaticity excitations caused by the $1/f$-type detuning noise [39] around the interdot transition region. Nevertheless, both scenarios predict an extrapolated fidelity approaching 100% for very fast ramps (orange dashed line in Fig. S6b), which is not observed here. Instead, our data is best described by accounting for an overall shift of $p$ by approximately 1.3%, independent of the ramp rate (yellow dotted line). This could indicate the presence of some source of error per transfer that is not caused by the time spent at the interdot transition region, besides not being effectively refocused by an echo sequence. Further investigations are needed to clarify what is the leading underlying mechanism.

In a longer chain of quantum dots, spin transport may be achieved by consecutive adiabatic tunnelling between nearest neighbours. The physical mechanisms expected to limit the transfer fidelity in a longer chain are largely present in the double dot system studied here. Extrapolating the observed coherence loss $p$ ~2% for a transfer between neighbouring sites would correspond to spin transfer across ~50 sites before the phase coherence decays to 1/e, or a distance ~2 $\mu$m (assuming a 40 nm site spacing). If only the spin polarization is needed e.g. for qubit readout [47], the electron could be transported over 2500 sites (or ~100 $\mu$m) before the polarization decays to 1/e for the spin-↑ case. While this accuracy of the spin transfer process indicates that coherent coupling between remote qubits is achievable, a fault-tolerant quantum computing architecture relying on qubit movement will require a device setup tailored to enhance the transfer fidelity. From our study, we can identify the following as desirable features: (i) the ability to electrostatically control the interdot tunnel rate [26,27] to guarantee adiabatic passage; (ii) a reduction in the difference of Larmor frequencies in neighbouring sites, achievable by controlling the spin-orbit coupling [44] or operating at lower magnetic fields [48]; and (iii) improvements in the fabrication process leading to less charge noise.

To conclude, we have demonstrated that a single electron spin can be coherently transported from one site to the next in an isotopically-enriched silicon quantum dot system. Our results show that the transfer process can be regarded as a unitary phase rotation gate with an average gate fidelity of $(99.36 \pm 0.05)$%. Beyond quantum-dot-based qubit implementations, coherent electron shuttling could also facilitate the scale-up of donor-based quantum computers [49], and enable the use of long-lived nuclear spin qubits – either in donors [50] or in isoelectronic atoms [37] – which can be faithfully entangled with the electron spin carrying the quantum information across long distances. Our results indicate the practical possibility of adopting non-local quantum gates in future error-corrected, quantum processors [4–7] based in silicon, and in the nearer term, will enable increased connectivity in few-qubit devices [8]. From the perspective of scalability, coherent spin transport could allow the spacing out of dense modules of physical qubits [9], addressing one of the most significant engineering challenges facing silicon-based quantum computing.

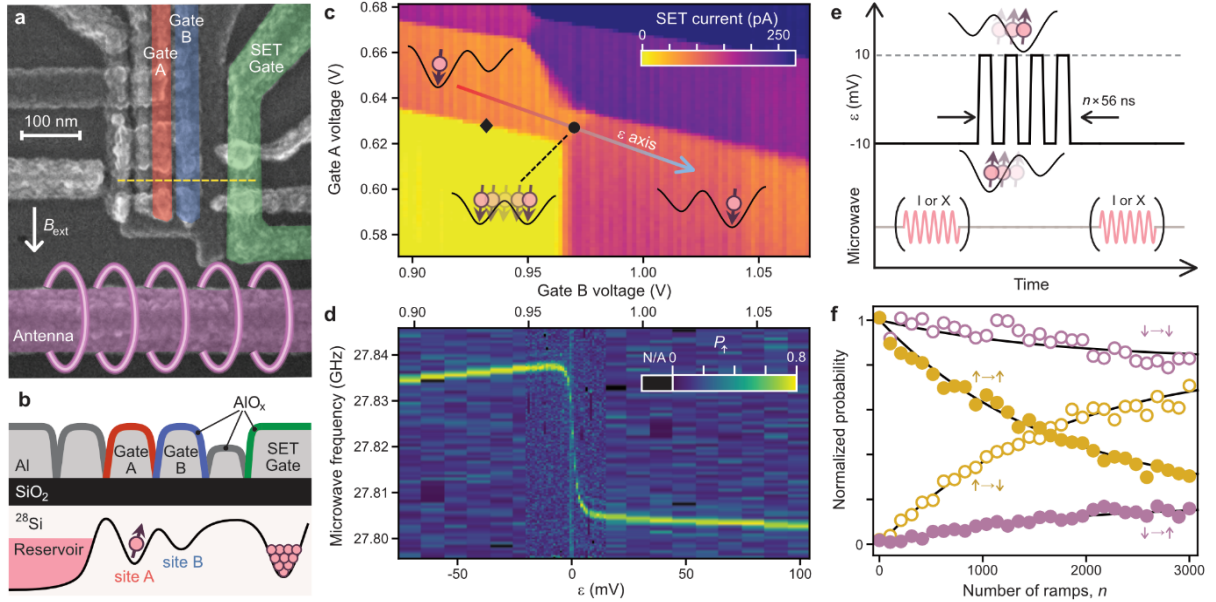

**Fig. 1. Qubit device and spin polarization transfer. (a)** False-colored scanning electron micrograph of an identical device. A linear array of quantum dots is formed in a silicon MOS structure underneath gates A (red) and B (blue). A single electron transistor (SET) under gate SET (green) is used for charge sensing. Spin control is performed by applying a microwave pulse to an on-chip ESR antenna (magenta). The arrow indicates the direction of an in-plane external magnetic field of 1 T (unless otherwise noted). **(b)** Cross-sectional schematic of the device. A single electron is loaded into quantum dot sites A and B and manipulated by gate voltage pulses applied on aluminum metal gates A and B. **(c)** Stability diagram and definition of the transfer axis. Charge configuration in the dot array is mapped through the SET current (a plane is subtracted). The arrow defines the gate-voltage axis used for qubit transport, ε. As ε is increased, the site where the electron resides changes from A to B. The interdot transition (ε = 0) is marked by a circle. Spin initialization and readout is performed at the diamond using spin-dependent tunneling to the reservoir (see Supplementary Information). **(d)** Gate voltage dependence of the qubit resonance frequency. The probability of detecting ↑ out of 100 events is measured after a 480 ns-long π pulse is applied. Data points with high reference signal (see Supplementary Information) are plotted in black. The rapid change at the interdot transition reveals a 30 MHz interdot resonance frequency separation. **(e)** Pulse schematic used for polarization transfer fidelity experiment. 368 ns-long π pulses are turned on (X) and off (I) to prepare both ↑ and ↓ initial states and to measure the probabilities of finding ↑ and ↓ states after the transfers. The total time in the ramp pulse section increases by 56 ns for each additional ramp. **(f)** Normalized probabilities of finding ↑ and ↓ states for the ↑ and ↓ inputs. The solid curves are fits to the data.

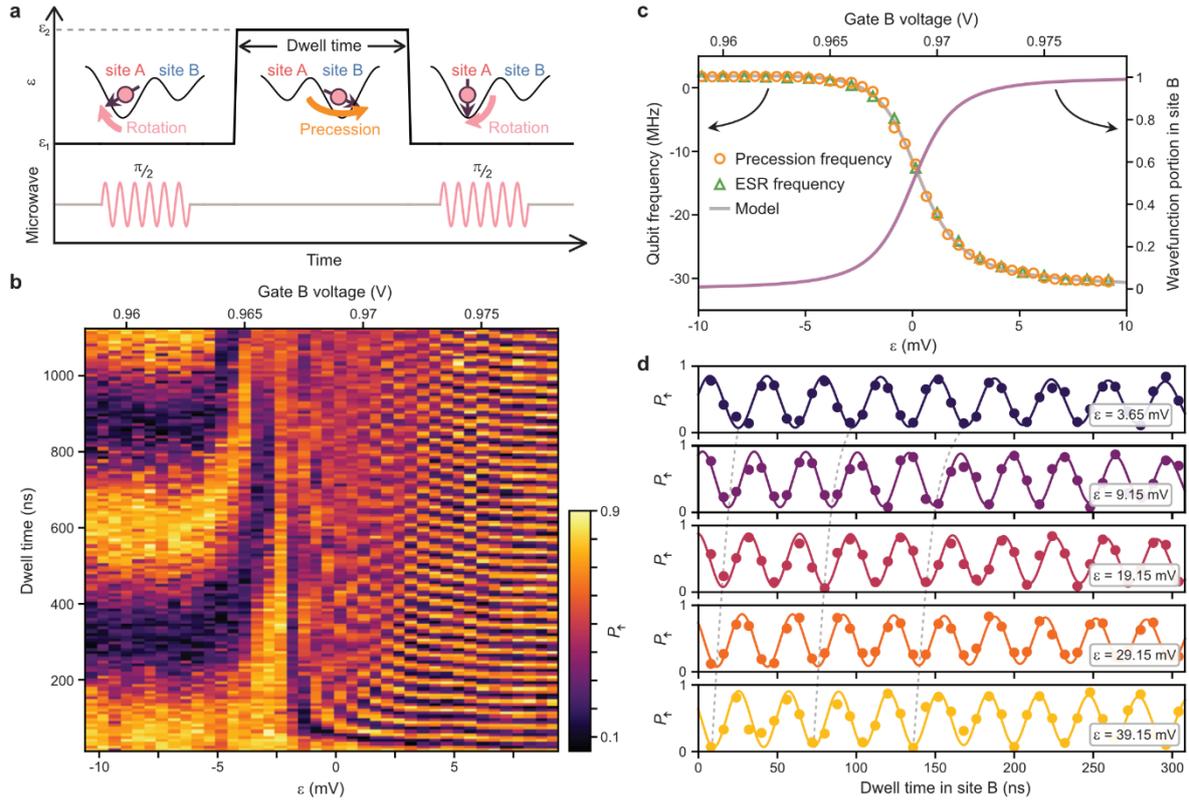

**Fig. 2. Coherent tunnelling spectroscopy.** (a) Pulse schematic used for the tunnelling spectroscopy. The spin, prepared in ↓ at site A, is first rotated to an equatorial state and then accumulates a phase during the interdot detuning pulse for a dwell time $t_{dwell}$, until a second $\pi/2$ pulse projects the phase to the polarization (↑ or ↓). (b) Tunnelling spectroscopy performed across the interdot transition. The continuous fringe evolution demonstrates the phase coherence during the tunnelling process. (c) Qubit spectrum extracted from the precession frequency (orange dots) as well as from the ESR spectrum (green triangles, offset by 27.8354128 GHz). The grey curve shows a fit to a four-level model with spin-dependent tunnel couplings. The purple solid line plots the spin-↓ electron wavefunction portion in site B calculated from the model (the one for the ↑ case overlaps with this). (d) Shuttling process as a phase gate. Rapid, 30 MHz phase rotations in the site B region are observed down to $t_{dwell} = 8$ ns. Dashed lines are guides to the eye for the first, third and fifth oscillation valleys.

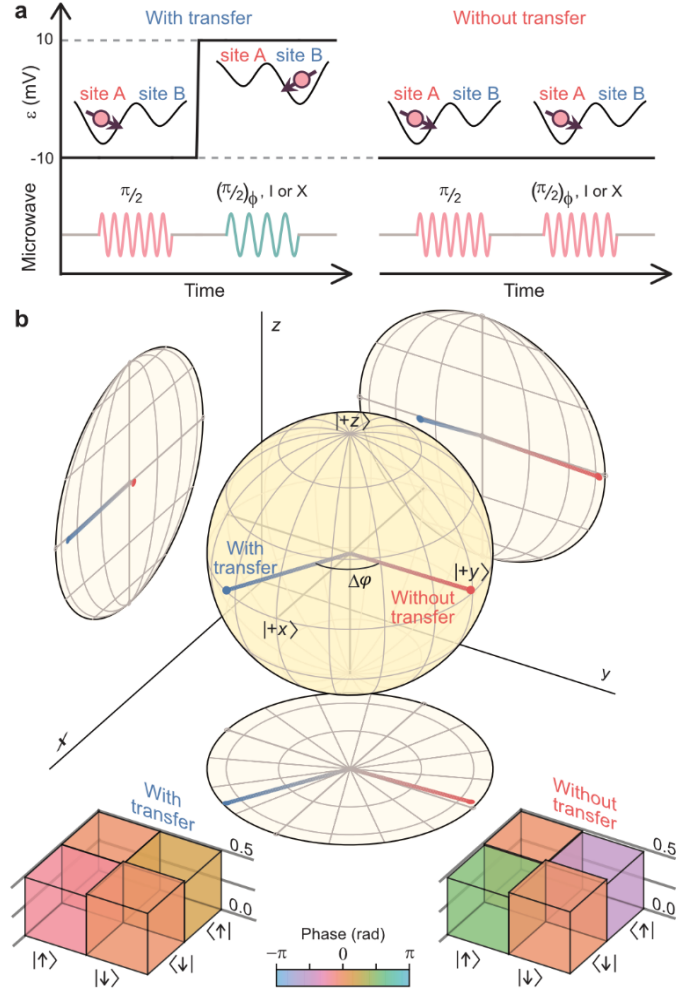

**Fig. 3. Quantum tomography of spin states with and without transfer. (a)** Schematic representation of the quantum state tomography experiments. $(\pi/2)_\phi$ denotes a $\pi/2$ ESR rotation along an axis within the *xy* plane whose azimuth angle from the *x* axis is $\phi$. The first eight multiples of $\pi/4$ are used for $\phi$. I and X represent an identity and a $\pi$ rotation along *x*, respectively. **(b)** Bloch sphere representation of reconstructed spin states before (red) and after (blue) an inter-site transfer process. The projections onto the *xy*- (bottom), *yz*- (right), and *zx*- (left) planes are also displayed. The primary net effect of the transfer process is the phase shift $\Delta\varphi$, rooted in the site-dependence of qubit frequency. The insets show the amplitude (height) and phase (colour) of the density matrix elements for individual states.

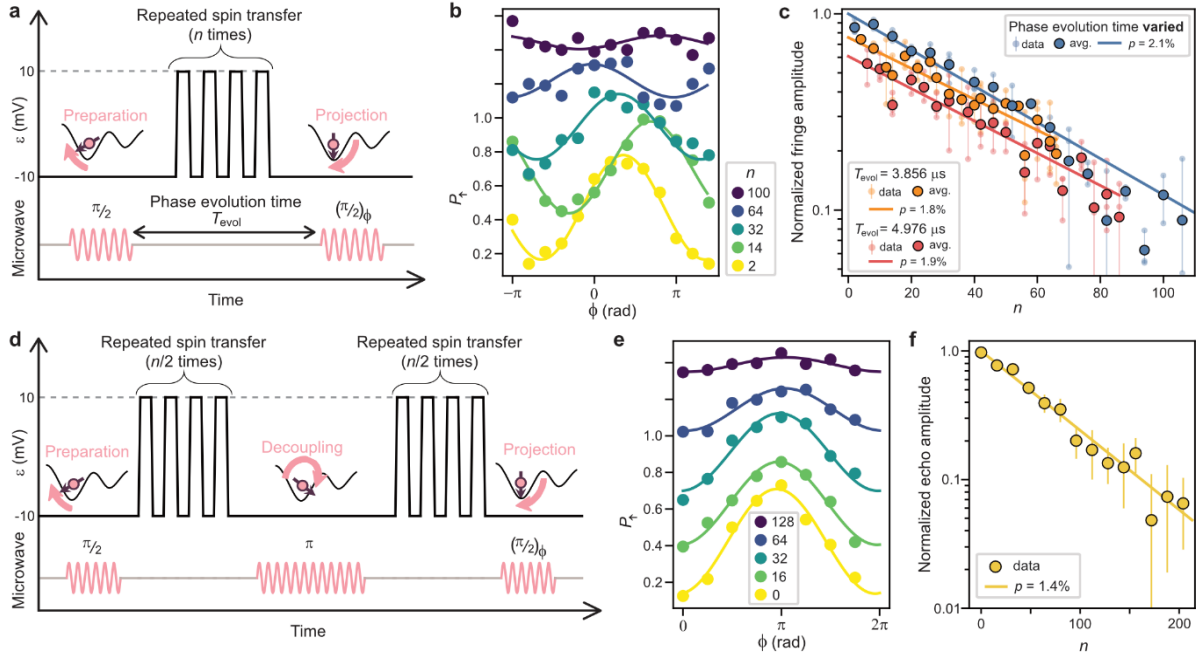

**Fig. 4. Coherent transfer fidelity characterization. (a)** Pulse sequence for shuttle fidelity characterization. Transport ramps are repeated a number ($n$) of times between the two $\pi/2$ pulses, whose interval is denoted as the phase evolution time, $T_{\text{evol}}$. **(b)** Fringes observed after shuttle ramps for various $n$ (with varying $T_{\text{evol}}$). Curves plot the fit results. Traces are offset for clarity. **(c)** Normalized fringe amplitudes as a function of $n$ with $T_{\text{evol}}$ varied (blue) or fixed (orange and red). Curves plot the fit results to an exponential decay. Smaller decay amplitudes for the data with $T_{\text{evol}}$ fixed are consistent with the expected reduction of coherence during $T_{\text{evol}}$. **(d)** Pulse sequence for echoed shuttle fidelity characterization. All microwave pules are applied at tone A and with the spin in site A. **(e)** Echo fringes observed for various $n$ along with the fit results. **(f)** Normalized echo amplitudes as a function of $n$ with a fit to an exponential decay. Error bars represent the $1\sigma$ confidence intervals of the echo amplitudes.

**References**

1. Campbell, E. T., Terhal, B. M., & Vuillot, C. Roads towards fault-tolerant universal quantum computation. *Nature* **549**, 172–179 (2017).

2. Jones, N. C. *et al*. Layered Architecture for Quantum Computing. *Phys. Rev. X* **2**, 031007 (2012). https://doi.org/10.1103/PhysRevX.2.031007

3. Fowler, A. G., Mariantoni, M., Martinis, J. M. & Cleland, A. N. Surface codes: Towards practical large-scale quantum computation. *Phys. Rev. A*, 86, 032324 (2012).

4. Jochym-O'Connor, T. & Bartlett, S. D. Stacked codes: Universal fault-tolerant quantum computation in a two-dimensional layout. *Phys. Rev. A* **93**, 022323 (2016).

5. Bombín, H. Gauge color codes: optimal transversal gates and gauge fixing in topological stabilizer codes. *New J. Phys.* **17**, 083002 (2015).

6. Krishna, A. & Poulin, D. Topological wormholes: Nonlocal defects on the toric code *Phys. Rev. Research* **2**, 023116 (2020).

7. Fawzi, O., Grospellier, A. & Leverrier, A. Constant Overhead Quantum Fault-Tolerance with Quantum Expander Codes, *Foundations of Computer Science (FOCS) 2018 IEEE 59th Annual Symposium on*, 743-754 (2018). https://doi.org/10.1109/FOCS.2018.00076.

8. Linke, N. *et al*. Experimental comparison of two quantum computing architectures. *Proc. Natl. Acad. Sci. USA* **114**, 3305-3310 (2017).

9. Vandersypen, L. M. K. *et al*. Interfacing spin qubits in quantum dots and donors - hot, dense and coherent. *npj Quantum Information*, 34, 1-10 (2017). https://doi.org/10.1038/s41534-017-0038-y

10. Taylor, J. *et al.* Fault-tolerant architecture for quantum computation using electrically controlled semiconductor spins. *Nat. Phys.* **1,** 177–183 (2005).

11. Li, R. *et al*. A Crossbar Network for Silicon Quantum Dot Qubits. *Sci. Adv.* **4**, eaar3960 (2018). https://advances.sciencemag.org/content/4/7/eaar3960

12. Yoneda, J. *et al.* A quantum-dot spin qubit with coherence limited by charge noise and fidelity higher than 99.9%. *Nat. Nanotechnol.* **13**,102–106 (2018).

13. Huang, W. *et al*. Fidelity benchmarks for two-qubit gates in silicon. *Nature* **569**, 532-536 (2019).

14. Zajac, D. M. *et al*. Resonantly driven CNOT gate for electron spins. *Science* **359**, 439-442 (2018).

15. Watson, T. F. *et al*. A programmable two-qubit quantum processor in silicon. *Nature* **555**, 633–637 (2018).

16. Yang, C. H. *et al*. Operation of a silicon quantum processor unit cell above one kelvin. *Nature* **580**, 350–354 (2020).

17. Petit, L. *et al*. Universal quantum logic in hot silicon qubits. *Nature* **580** (7803), 355–359 (2020).

18. Yoneda, J. *et al*. Quantum non-demolition readout of an electron spin in silicon. *Nat. Commun.* **11**, 1144 (2020). https://doi.org/10.1038/s41467-020-14818-8

19. Veldhorst, M. *et al*. Silicon CMOS architecture for a spin-based quantum computer. *Nat. Commun.* **8,** 1766 (2017). https://doi.org/10.1038/s41467-017-01905-6

20. Pillarisetty, R. *et al*. Qubit Device Integration Using Advanced Semiconductor Manufacturing Process Technology. *IEDM* 6.3.1-6.3.4 (2018). https://doi.org/10.1109/IEDM.2018.8614624

21. Vinet, M. *et al*. Towards scalable silicon quantum computing. *IEDM* 6.5.1-6.5.4 (2018). https://doi.org/10.1109/IEDM.2018.8614675.

22. Franke, D. P., Clarke, J. S., Vandersypen, L. M. K., & Veldhorst, M. Rent's rule and extensibility in quantum computing. *Microprocessors and Microsystems* **67**, 1–7 (2019).

23. DiVincenzo, D. The physical implementation of quantum computation. *Fort. Phys.* **48**, 771–783 (2000).

24. Buonacorsi, B. *et al*. Network architecture for a topological quantum computer in silicon. *Quantum Sci. and Technol.* **4**, 025003 (2019).

25. Kandel, Y. P., Qiao, H., Fallahi, S., Gardner, G. C., Manfra, M. J. & Nichol, J. M. Coherent spin state transfer via Heisenberg exchange. *Nature* **573**, 553–557 (2019).

26. Sigillito, A. J., Gullans, M. J., Edge, L. F., Borselli, M. & Petta, J. R. Coherent transfer of quantum information in a silicon double quantum dot using resonant SWAP gates. *npj Quantum Information*, 5, 1–7 (2019). https://doi.org/10.1038/s41534-019-0225-0

27. Takeda, K., Noiri, A., Yoneda, J., Nakajima, T. & Tarucha, S. Resonantly Driven Singlet-Triplet Spin Qubit in Silicon. *Phys. Rev. Lett.* **124**, 117701 (2020).

28. Samkharadze, N. *et al*. Strong spin-photon coupling in silicon. *Science* **359**, 1123-1127 (2018).

29. Landig, A. J. *et al*. Virtual-photon-mediated spin-qubit–transmon coupling, *Nat. Commun*. **10**, 5037 (2019). https://doi.org/10.1038/s41467-019-13000-z

30. Borjans, F., Croot, X. G., Mi, X., Gullans, M. J. & Petta, J. R. Resonant microwave-mediated interactions between distant electron spins. *Nature* **577**, 195–199 (2020).

31. Skinner, A. J., Davenport, M. E. & Kane, B. E. Hydrogenic Spin Quantum Computing in Silicon: A Digital Approach. *Phys. Rev. Lett.* **90**, 087901 (2003).

32. Greentree, A. D., Cole, J. H., Hamilton, A. R. & Hollenberg, L. C. L. Coherent electronic transfer in quantum dot systems using adiabatic passage. *Phys. Rev. B* **70**, 235317 (2004).

33. Bertrand, B. *et al*. Fast spin information transfer between distant quantum dots using individual electrons, *Nat. Nanotechnol*. **11**, 672–676 (2016).

34. Fujita, T., Baart, T. A., Reichl, C., Wegscheider, W. & Vandersypen, L. M. K. Coherent shuttle of electron-spin states. *npj Quantum Information*, 22, 1–5, (2017). https://doi.org/10.1038/s41534-017-0024-4

35. Nakajima, T. *et al*. Coherent transfer of electron spin correlations assisted by dephasing noise. *Nat. Commun.* **9**, 2133 (2018). https://doi.org/10.1038/s41467-018-04544-7

36. Mills, A. *et al*. Shuttling a single charge across a one-dimensional array of silicon quantum dots. *Nat. Commun.* **10**, 1063 (2019). https://doi.org/10.1038/s41467-019-08970-z

37. Hensen, B. *et al*. A silicon quantum-dot-coupled nuclear spin qubit. *Nat. Nanotechnol.* **15**, 13–17 (2020).

38. Huang, P. and Hu, X. Spin qubit relaxation in a moving quantum dot. *Phys. Rev.* B **88**, 075301 (2013).

39. Li, X., Barnes, E., Kestner, J. P., & Das Sarma, S. Intrinsic errors in transporting a single-spin qubit through a double quantum dot. *Phys. Rev. A* **96**, 012309 (2017).

40. Krzywda, J. A., & Cywiński, Ł. Adiabatic electron charge transfer between two quantum dots in presence of 1/$f$ noise. *Phys. Rev. B* **101**, 035303 (2020).

41. Buonacorsi, B., Shaw, B. & Baugh, J. Simulated coherent electron shuttling in silicon quantum dots. arXiv:2003.08018 (2020).

42. Ginzel, F., Mills, A. R., Petta, J. R. & Burkard, G. Spin shuttling in a silicon double quantum dot. arXiv:2007.03598v1 (2020).

43. Stano, P. & Fabian, J. Spin-orbit effects in single-electron states in coupled quantum dots. *Phys*. *Rev*. *B* **72**, 155410 (2005).

44. Tanttu, T. *et al*. Controlling Spin-Orbit Interactions in Silicon Quantum Dots Using Magnetic Field Direction. *Phys*. *Rev*. *X* **9**, 021028 (2019). https://doi.org/10.1103/PhysRevX.9.021028

45. Laucht, A. *et al*. Electrically controlling single-spin qubits in a continuous microwave field. *Sci*. *Adv*. **1**, e1500022 (2015). https://doi.org/10.1126/sciadv.1500022

46. Biercuk, M. J., Doherty, A. C. & Uys, H. Dynamical decoupling sequence construction as a filter-design problem. *J. Phys. B* **44**, 154002 (2011).

47. Baart, T. A., Shafiei, M., Fujita, T., Reichl, C., Wegscheider, W. & Vandersypen, L. M. K. Single-spin CCD. *Nat. Nanotechnol.* **11**, 330–334 (2016).

48. Zhao, R. *et al*. Single-spin qubits in isotopically enriched silicon at low magnetic field. *Nat. Commun.* **10**, 5500 (2019). https://doi.org/10.1038/s41467-019-13416-7

49. Pica. G., Lovett, B. W., Bhatt, R. N., Schenkel, T. & Lyon, S. A. Surface code architecture for donors and dots in silicon with imprecise and nonuniform qubit couplings. *Phys. Rev. B* **93**, 035306 (2016).

50. Dehollain, J. P. *et al*. Bell's inequality violation with spins in silicon. *Nat. Nanotechnol.*, **11**, 242–246 (2016).

51. Itoh, K. M. and Watanabe, H. Isotope engineering of silicon and diamond for quantum computing and sensing applications. *MRS Commun*. **4**, 143-157 (2014).

52. Poyatos, J. F., Cirac, J. I., and Zoller, P. Complete Characterization of a Quantum Process: The Two-Bit Quantum Gate. *Phys. Rev. Lett.* **78**, 390-393 (1997).

53. Nielsen, M. A. A simple formula for the average gate fidelity of a quantum dynamical operation. *Phys. Lett.* A **303**, 249-252 (2002).

**Acknowledgements** We thank C. Escott for discussions and assistance with the work. We acknowledge support from the Australian Research Council (FL190100167 and CE170100012), the US Army Research Office (W911NF-17-1-0198), and the NSW Node of the Australian National Fabrication Facility. The views and conclusions contained in this document are those of the authors and should not be interpreted as representing the official policies, either expressed or implied, of the Army Research Office or the US Government. K.M.I. acknowledges support from a Grant-in-Aid for Scientific Research by MEXT.

**Author contributions** J.Y. and W.H. performed the experiments and analysed the data. J.Y., W.H., M.F. and A.S. discussed the results. C.H.Y., T.T., W.G. and R.C.C.L. contributed to the measurement setup. K.W.C. and F.E.H. fabricated the device. K.M.I. prepared and supplied the [28]Si wafer. S.D.B. contributed to the interpretation of the results. J.Y. and A.S. wrote the manuscript with input from all co-authors. A.M., A.L. and A.S.D supervised the project.

**Author information** Correspondence and requests for materials should be addressed to J.Y. or A.S.D.

## Supplementary Information

### Measurement setup

The device is an isotopically enriched (residual $^{29}$Si concentration of 800 ppm) silicon MOS quantum dot system [51] as reported in Ref. [13]. The experiment was performed in an Oxford Instruments Kelvinox dilution refrigerator. A 4-channel arbitrary waveform generator (Lecroy Arbstudio 1104), which is triggered by a TLL pulse generator (SpinCore PulseBlaster-ESR), is used to generate two-channel gate pulses (applied to gates A and B) as well as to provide the digital modulation signals to shape ESR microwave pulses through external in-phase/quadrature modulation ports of a Keysight 8267D microwave source. We found that the result can be impacted by the jitter between the gate pulses and the microwave pulses when they were sourced from different instruments at early stages of our experiments. All data except Fig. 1e were acquired with the ESR frequency feedback protocol [13]. The SET current is amplified by a room temperature I/V converter (Femto DLPCA-200), filtered at 10 kHz using an 8-order Bessel filter and sampled by an oscilloscope (pico Technology PicoScope 4824).

### Measurement pulse

The spin is first initialized by selectively loading a spin-↓ electron from the reservoir to site A by aligning the reservoir Fermi energy between the spin Zeeman sublevels at the diamond marker in Fig. 1c. Then the gate voltage is set to a point on the $\varepsilon$ axis. Over the course of the experiments, a drift in the gate voltage is occasionally observed, which we compensate by redefining the origin of the $\varepsilon$ axis. After completing the transport ramps along the $\varepsilon$ axis, the gate voltage is configured to the same position as used for initialization to perform readout based on the spin-selective tunnelling from site A to the reservoir. The tunnelling causes a blip in the SET current as site A charge state changes from filled to empty, which we interpret as a spin-↑ event. We also record the SET current after a sufficiently long time (compared with the ~1 ms tunnelling time) as a reference signal. The probability that this reference signal is at the empty level will be low but finite due to non-unity visibility if the readout level is properly aligned.

### Polarization fidelity analysis

The spin-dependent polarization transfer fidelities, $F_{\text{pol}}^{\uparrow}$ and $F_{\text{pol}}^{\downarrow}$, are obtained from the probabilities of finding the same (or opposite) spin state as the input state after *n* consecutive transfer ramps, $F_{\text{pol}}^{\uparrow,n}$ (or $1 - F_{\text{pol}}^{\uparrow,n}$) and $F_{\text{pol}}^{\downarrow,n}$ (or $1 - F_{\text{pol}}^{\downarrow,n}$). We model these probabilities as

$$\begin{pmatrix} F_{\text{pol}}^{\uparrow,n} & 1 - F_{\text{pol}}^{\downarrow,n} \\ 1 - F_{\text{pol}}^{\uparrow,n} & F_{\text{pol}}^{\downarrow,n} \end{pmatrix} = \begin{pmatrix} F_{\text{pol}}^{\uparrow} & 1 - F_{\text{pol}}^{\downarrow} \\ 1 - F_{\text{pol}}^{\uparrow} & F_{\text{pol}}^{\downarrow} \end{pmatrix}^n,$$

treating the transfer-induced spin flip as a memory-less process. This formula is found to explain the observed $n$ dependence well, assuming a common visibility pre-factor and no error in the $\pi$ rotation(s). We fit the four probability traces simultaneously using this expression, and extract the values of $F_{\text{pol}}^{\uparrow}$ and $F_{\text{pol}}^{\downarrow}$.

**Double-dot spin tunnelling model**

The Ramsey-type spectroscopy measures the energy splitting between the instantaneous eigenstates. The observed spectrum can be well explained considering a model with a single orbital in each quantum dot, without taking into account the intradot valley and orbital excitations [41]. The most general model could contain spin effects both in the spin-conserving tunnelling (in the form of a spin dependence of the coupling) as well as a spin-flip tunnelling term. Both of these effects may occur as a combination of the effects of the kinetic momentum of the electron leading to some spin-orbit field, as well as a small difference in the quantisation axes of the dots, due to the variability in *g*-tensors [44]. For the particular purpose of describing the Ramsey spectroscopy data, we can neglect the contribution from a small spin-flipping tunnelling term (which has reduced impact on the energy splitting, generating effectively a transverse field). Then we can treat the state hybridization separately for individual spin orientations (↑ and ↓) – see Fig. S2a and b. This simple four-level model predicts the qubit frequency $f_Q$ to be

$$f_Q = \frac{f_A + f_B}{2} + \frac{1}{2}\sqrt{\left(\alpha\varepsilon - \frac{f_A - f_B}{2}\right)^2 + (t_c - t_s)^2} - \frac{1}{2}\sqrt{\left(\alpha\varepsilon + \frac{f_A - f_B}{2}\right)^2 + (t_c + t_s)^2}$$

where $f_{A(B)}$ is the bare qubit frequency at site A (B). Here $\alpha$ denotes the effective leverarm of the gate B voltage change along the $\varepsilon$ axis on the energy difference between the localized states, $t_c$ the tunnel coupling and $t_s$ its spin dependence due to spin-orbit coupling (positive if it is larger for spin-↑). $f_{A(B)}$ is further parametrized as $f_{A(B)} = f_Z + \eta_{A(B)}\varepsilon + (-)\Delta f_{AB}/2$, where $f_Z$ is the average of bare qubit frequencies at $\varepsilon = 0$, $\eta_{A(B)}$ accounts for the Stark shift constant and $\Delta f_{AB}$ gives the qubit frequency difference between sites at $\varepsilon = 0$. We find that this fully explains the qubit frequency $f_Q$ measured along the $\varepsilon$ axis over 200 mV (Fig. S2c). We note that the origin of $\varepsilon$ is simultaneously determined from this modelling. Using the leverarm extracted from a separate experiment (0.21 eV/V), the best fit is obtained for $t_c$ = 104 GHz, $t_s$ = -3.4 MHz, $\eta_A$ = 39 MHz/V, $\eta_B$ = -7.1 MHz/V and $\Delta f_{AB}$ = 33.4 MHz.

**State tomography**

The pre-transfer electron spin state is prepared to $|+y\rangle$ by a π/2 ESR pulse in site A after initialization to the ↓ state. The spin is then either transferred to site B or kept at site A, before we perform a pre-measurement control. Ten kinds of pre-measurement controls – eight π/2 rotations with varying phases (controlled through the microwave phase ϕ), as well as identity (I) and π-rotation (X) operations – are used to effectively change the measurement basis state $|\psi_v\rangle$ of the readout of ↑ which follows. An overcomplete number of π/2 rotation axes are employed to help reduce the measurement bias error. In addition, the state-preparation and measurement fidelity $F_{M,\uparrow(\downarrow)}$ is obtained by interleaved measurement of the ↑ probabilities with the spin prepared in ↑ or ↓. $F_{M,\uparrow(\downarrow)}$ is measured to be 80.4% (87.9%), allowing for the measurement visibility correction.

The density matrix of the pre- or post-transfer spin state, $\rho$, is then reconstructed from the corrected ↑ probabilities, $p_v$, after 4,000 repetitions for each of ten measurement basis states, using maximum likelihood estimation. We restrict $\rho$ to be non-negative Hermitian and unit trace by expressing it through a complex matrix, $L$:

$$\rho(\boldsymbol{\ell}) = \frac{L^\dagger L}{\text{tr}(L^\dagger L)}.$$

$L$ is a 2 × 2 lower triangular matrix whose diagonal elements are real, and has three independent parameters, denoted by $\boldsymbol{\ell} = (\ell_1, \ell_2, \ell_3)$. To obtain the closest physical $\rho$, the following cost function, $C$, is minimized:

$$C(\boldsymbol{\ell}) = \sum_{v=1}^{10} \frac{(\langle\psi_v|\rho(\boldsymbol{\ell})|\psi_v\rangle - p_v)^2}{2\langle\psi_v|\rho(\boldsymbol{\ell})|\psi_v\rangle}.$$

The state fidelity of the resulting $\rho$ is defined by $\left(\text{Tr}\left[\sqrt{\sqrt{\rho_{\text{ideal}}}\rho\sqrt{\rho_{\text{ideal}}}}\right]\right)^2$, where $\rho_{\text{ideal}}$ is the density matrix of the closest pure state on the equator of the Bloch sphere. That is, the ideal transfer process is considered as a phase gate which does not alter the spin polarization. We note that by comparing to the closest equatorial pure state in this way, we implicitly ignore any coherent phase error. To estimate the statistical error, a Monte Carlo simulation is performed to yield a distribution of the estimated state fidelities, from which the 1σ (68.27%) confidence intervals are calculated around its median value.

**Average coherent transfer fidelity**

We use the single-qubit average gate fidelity [52,53] as a measure of the faithfulness of the qubit transfer process. It is commonly used to quantify the fidelity of an operation and is defined by the average state fidelity of the output qubit state $\rho$ with respect to the output $\rho_{\text{ideal}}$ from the ideal gate, $\left(\text{Tr}\left[\sqrt{\sqrt{\rho_{\text{ideal}}}\rho\sqrt{\rho_{\text{ideal}}}}\right]\right)^2$, over all pure input states. Using the spin-dependent polarization transfer infidelities, $r^\uparrow = 1 - F_{\text{pol}}{}^\uparrow$ and $r^\downarrow = 1 - F_{\text{pol}}{}^\downarrow$, as well as the coherence loss per transfer $p$, the spin density matrix after a transfer $\rho$ can be expressed as

$$\rho = \mathcal{M}(\rho_{\text{ideal}}) = \begin{pmatrix} (1-r^\uparrow)\rho_{\text{ideal},00} + r^\downarrow \rho_{\text{ideal},11} & (1-p)\rho_{\text{ideal},01} \\ (1-p)\rho_{\text{ideal},10} & r^\uparrow \rho_{\text{ideal},00} + (1-r^\downarrow)\rho_{\text{ideal},11} \end{pmatrix}$$

where $\rho_{\text{ideal},ij}$ denotes the corresponding matrix element of $\rho_{\text{ideal}}$. $\mathcal{M}$ is a completely positive trace-preserving map describing the error associated with the transfer process. It is instructive to consider $\mathcal{M}$ as a cascade of dephasing and polarization-changing channels. When we model the polarization-changing channel in the operator-sum formalism through the Kraus operators $J_1 = \sqrt{1-r^\uparrow}\begin{pmatrix} 1 & 0 \\ 0 & 0 \end{pmatrix} + \sqrt{1-r^\downarrow}\begin{pmatrix} 0 & 0 \\ 0 & 1 \end{pmatrix}$, $J_2 = \sqrt{r^\uparrow}\begin{pmatrix} 0 & 0 \\ 1 & 0 \end{pmatrix}$ and $J_3 = \sqrt{r^\downarrow}\begin{pmatrix} 0 & 1 \\ 0 & 0 \end{pmatrix}$, and the dephasing channel through $K_1 = \sqrt{1-\frac{p'}{2}}\begin{pmatrix} 1 & 0 \\ 0 & 1 \end{pmatrix}$ and $K_2 = \sqrt{\frac{p'}{2}}\begin{pmatrix} 1 & 0 \\ 0 & -1 \end{pmatrix}$, $\mathcal{M}$ can be represented by the six Kraus operators $K_1 J_1$, $K_1 J_2$, $K_1 J_3$, $K_2 J_1$, $K_2 J_2$ and $K_2 J_3$. The dephasing parameter $p'$ of the dephasing channel and the coherence loss per transfer $p$ are then related by $1 - p' = (1-p)/\sqrt{(1-r^\uparrow)(1-r^\downarrow)}$. The fidelity of $\rho$ with respect to $\rho_{\text{ideal}}$ depends on the input state. As an illustration, for equatorial states it will be $1 - \frac{p}{2}$, whereas for polarized states it is given by $1 - r^\uparrow (= F_{\text{pol}}{}^\uparrow)$ or $1 - r^\downarrow (= F_{\text{pol}}{}^\downarrow)$. By calculating its average over all pure input states, we obtain the fidelity of the transfer process as $1 - (r^\uparrow + r^\downarrow + 2p)/6$.

**Transport verification**

Here we provide experimental evidence that the qubit is indeed transported between the two sites every time the detuning is ramped across the zero detuning point in the fidelity characterization experiments. We note that while it is possible to simulate the expected diabatic tunneling probability within the model using the extracted Hamiltonian parameters and the detuning ramp rate, the existence of noise and excited levels may introduce a subtlety to the dynamics in the experiment, in which case the wavefunction (or position) of the electron may not adiabatically follow the ground state of the electric potential controlled via gate voltages (on gates A and B with ε typically swept between +/-10 mV). Furthermore, the actual gate-voltage pulse shapes at the device end may well deviate from the voltage pulses generated in the room temperature circuit, as they are sent through filtered coaxial cables inside the dilution refrigerator (designed for a bandwidth of 80 MHz in our setup). We find it impracticable to estimate the distorted pulse shape in the presence of the transmission non-idealities, such as the standing wave modes, non-linear phase responses and nanosecond-scale inter-gate skews, inside the cryostat. Therefore, we design the following control experiment to verify that the qubit alternates between sites A and B for the same number of times as the number of applied transfer ramps, especially when we use the same pulse shape as in the fidelity characterization experiments.

The central idea of our control experiment is that the spin phase acquired under detuning pulses is very sensitive to the time the qubit spends in each site, due to the site-dependence of the qubit frequency (given by the slightly different $g$-factors, see the main text). We can precisely determine the spin precession rate for a given charge configuration and detuning from the tunneling spectroscopy result (see Fig. 2c and Fig. S2c). When $\varepsilon$ is swept between +/-10 mV, the difference in the phase precession rate is 32.4 MHz if the qubit switches its site following the (orbital) ground state as expected, with the dominant contribution coming from $\Delta f_{AB}$. In the meantime, if the qubit somehow remains in the same site, the precession rate would change by at most 1.2 MHz, since the Stark shift and the tunneling hybridization effect for these values of $\varepsilon$ are much smaller. This means that during a 56 ns interval between the transfer ramps, the amount of spin phase accumulation will be around $3.6\pi$ in the case of adiabatic transfer as opposed to $0.13\pi$ in the absence of electron transfer.

In the actual experiment, we employ pulses with different odd numbers of ramps, $n = 1, 3$ and 5 (see Fig. S1a) and compare the qubit phases after the pulses. We henceforth denote the difference in the post-transfer spin phase between the pulses with the odd numbers of ramps, $n = i$ and $i+2$, by $\delta\phi_{i+2,i}$. The phase difference $\delta\phi_{i+2,i}$ will be approximately $3.6\pi$ when the electron does return to site A and spends roughly 56 ns longer time there (or, equivalently, 56 ns shorter time in site B) after the $i$-th ramp is completed. Conversely, if the electron fails to change its site, $\delta\phi_{i+2,i}$ will be much smaller ($\sim 0.13\pi$).

We measure the fringes after the detuning ramps and quantify $\delta\phi_{i+2,i}$ from the difference in the fringe phase as a function of $\varepsilon_1$, where $\varepsilon_1$ is the specified value of $\varepsilon$ at the ramp starting point (see Fig. S1a). By changing $\varepsilon_1$ in sufficiently small steps, this protocol allows us to evaluate $\delta\phi_{i+2,i}$ larger than $2\pi$ experimentally (with no ambiguity given $\delta\phi_{i+2,i} \approx 0$ for $\varepsilon_1 = +10$ mV). It is important that the microwave phases of the two ESR pulses in this Ramsey-type sequence (applied at different sites and thus with different tones) be defined consistently for different numbers of ramps. The fringes obtained for various values of $\varepsilon_1$ are exemplified in Fig. S1b-f. Figure S1g plots the extracted values of $\delta\phi_{3,1}$ and $\delta\phi_{5,3}$ as a function of $\varepsilon_1$. The two traces agree well, suggesting the reliability of the phase measurement protocol and the consistency in the trajectories as $n$ is incremented. When $\varepsilon_1$ is large (e.g. point f), the phase difference is constantly small as the electron is not shuttled between sites and stays in site B. At around $\varepsilon_1 = 2$ mV (point e), the spin phase starts to pick up the tunneling hybridization effect around the anticrossing. Around $\varepsilon_1$ = -10 mV (point b), $\delta\phi_{i+2,i}$ reaches 3.6π, which can only be accounted for by a decrease in time spent in site B by ~56 ns upon increasing *n* by 2, indicating that following the *i*-th ramp, the qubit moves from site B to A (the (*i*+1)-th ramp) and, after dwelling for ~56 ns in site A, goes back to site B (the (*i*+2)-th ramp). We therefore conclude that for $\varepsilon_1$ = -10 mV, the same value that was used for the transfer fidelity measurements, the electron consistently moves between the sites for *n* times.

The data is even more compelling when we compare them with the calculation based on the qubit spectrum. The traces are already well explained by assuming that the actual qubit detuning is identical to the specified detuning pulse and that the dynamics is completely adiabatic (see the grey solid curve in Fig. S1g). For illustration purposes, we plot the results for two other transfer functions between the specified detuning pulse and the actual qubit detuning, which yield a slightly better alignment with the data, compared to the unfiltered case: a Butterworth filter (6$^{th}$-order 40 MHz lowpass) and a Chebyshev filter (type 1, 1$^{st}$-order bandstop with a 25-50 MHz stopband and a 1dB ripple level). The trajectories of the qubit detuning and frequency for $\varepsilon_1$ = -10 mV are also shown for individual cases in Fig. S1h-m. In all cases, the simulations verify that the qubit is transported between sites, further reinforcing our conclusion.

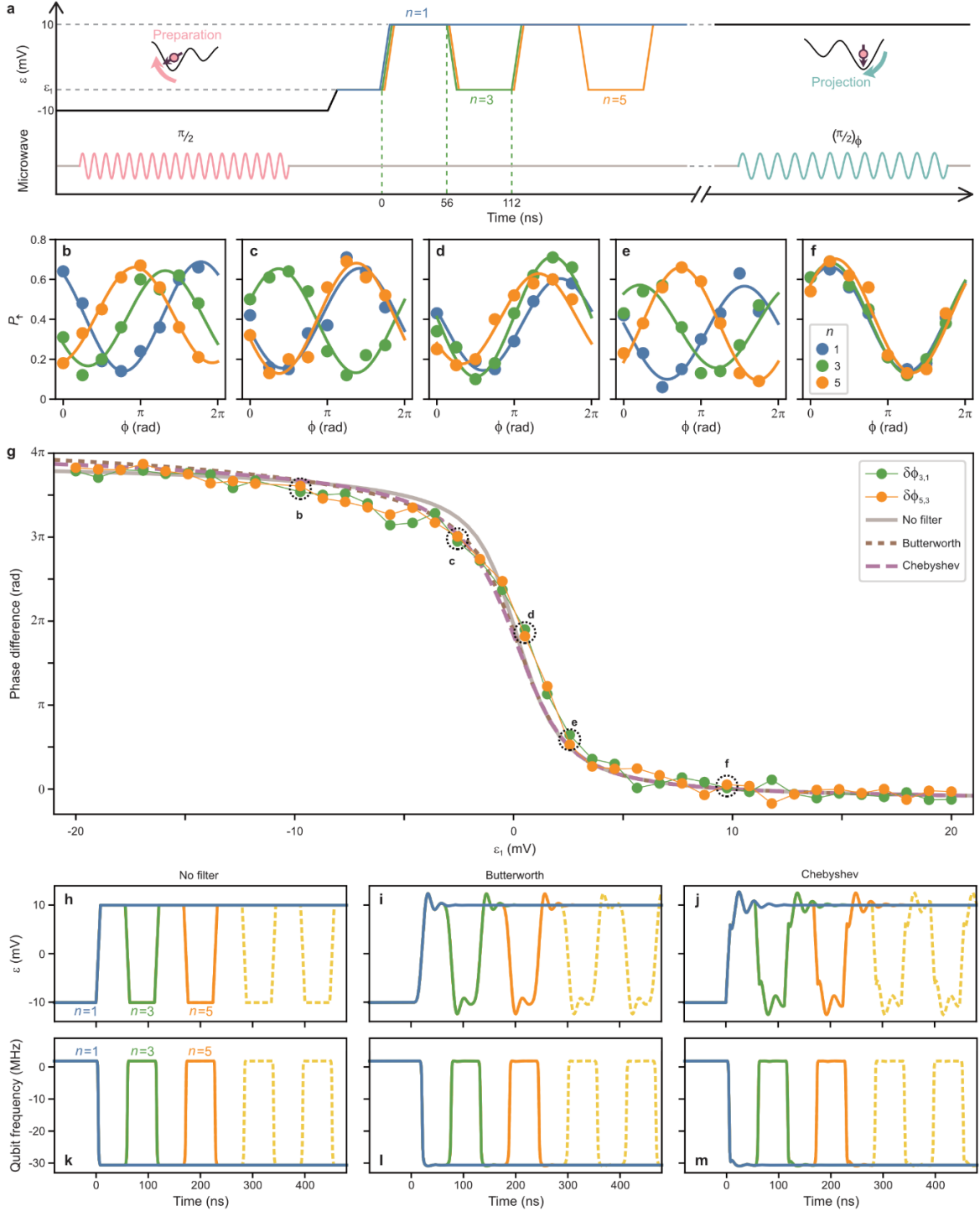

**Fig. S1. Transport verification. (a)** Pulse sequences used for transport verification. Laterally offset for clarity. **(b)-(f)** Ramsey fringes observed for various $\varepsilon_1$ (specified in (g)) along with fit curves from which to extract the phase. **(g)** Measured and calculated phase differences $\delta\phi_{i+2,i}$ as a function of $\varepsilon_1$. For illustration, we employ three transfer functions between the specified detuning pulse and the actual qubit detuning. Data points with small (<0.04) fringe amplitudes are omitted. **(h)-(j)** Qubit detuning trajectories for exemplified transfer functions. **(k)-(m)** Corresponding trajectories of the qubit frequency.

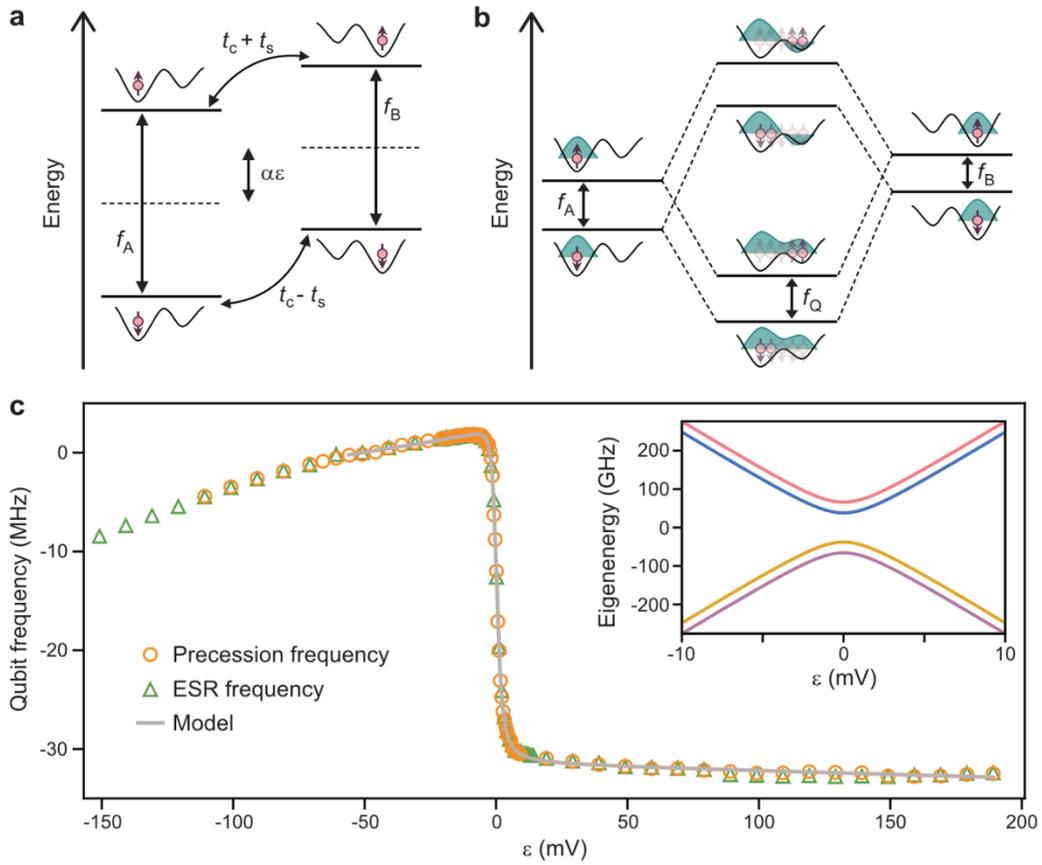

**Fig. S2. Qubit spectrum modelling.** **(a)** Illustration of the Hamiltonian terms used in the model. **(b)** Energy diagram showing the tunnelling hybridization. The left and rightmost columns depict the bare energy levels in individual sites in the absence of tunnelling. Energy levels in the middle column result from tunnelling hybridization. $f_Q$ gives the qubit frequency. **(c)** Wide-span qubit spectrum and the fit to the model. The finite slopes in the far detuned region are due to the Stark shift. The inset shows the eigen-energies calculated from the model.

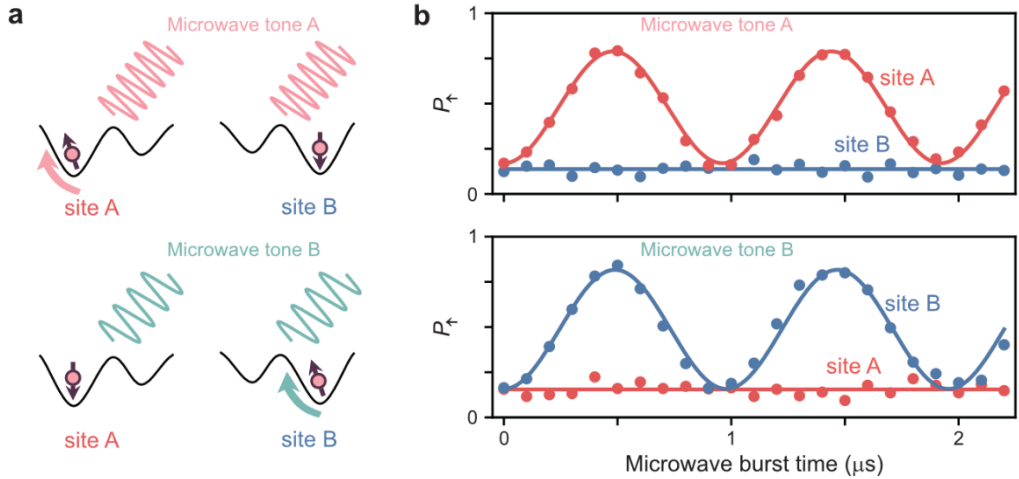

**Fig. S3. Site-dependent qubit response to a fixed ESR control tone. (a)** Schematics showing the site-dependent qubit response to ESR microwave pulses. The qubit rotates only when its site matches the microwave tone used. **(b)** Rabi oscillations observed by turning on and off the ESR drive by switching the qubit site.

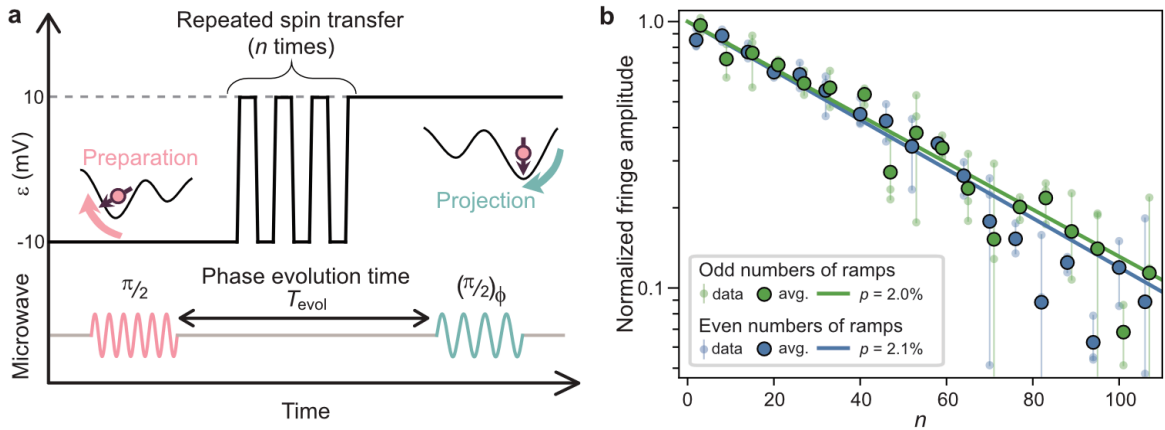

**Fig. S4. Coherent transfer fidelity measurement with odd numbers of ramps. (a)** Pulse schematics for odd numbers of ramps. The projection ESR pulse is applied while the qubit is in site B. **(b)** Comparison of the fringe amplitude decays for the cases of odd (green) and even $n$ (blue, same data as in Fig. 4c). The phase evolution time is varied in both cases.

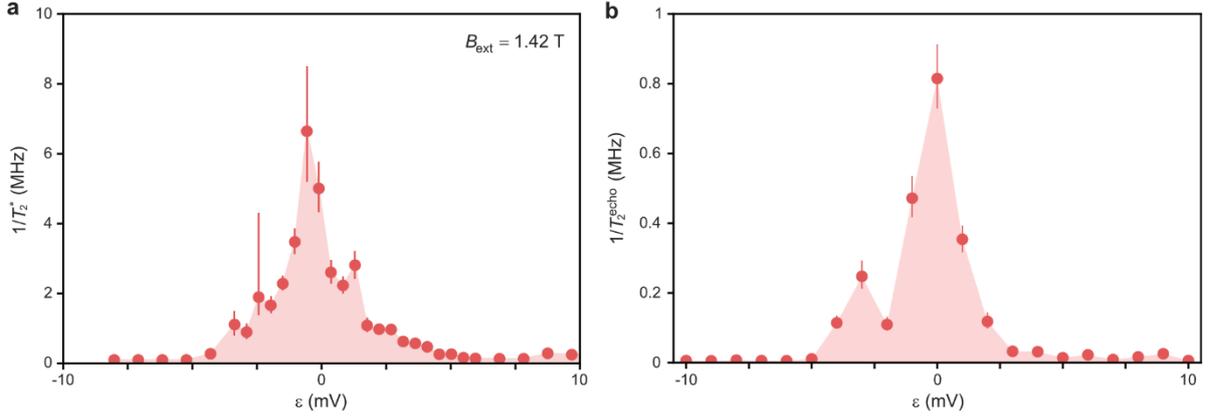

**Fig. S5. Coherence times measured near zero detuning. (a)** The inverse of $T_2^*$ plotted as a function of ε. In this data set, the external field applied (1.42 T) is larger by a factor of 1.42, and $\Delta f_{AB}$ is enhanced to 49.4 MHz. The qubit is expected to be proportionally more susceptible to charge noise around ε = 0. The device gate-voltage configuration is slightly different as well. **(b)** The inverse of Hahn echo time $T_2^{echo}$ as a function of ε. The data shows that dephasing around ε = 0 is mostly refocused by an echo sequence (note the 10-fold difference in the y-scale).

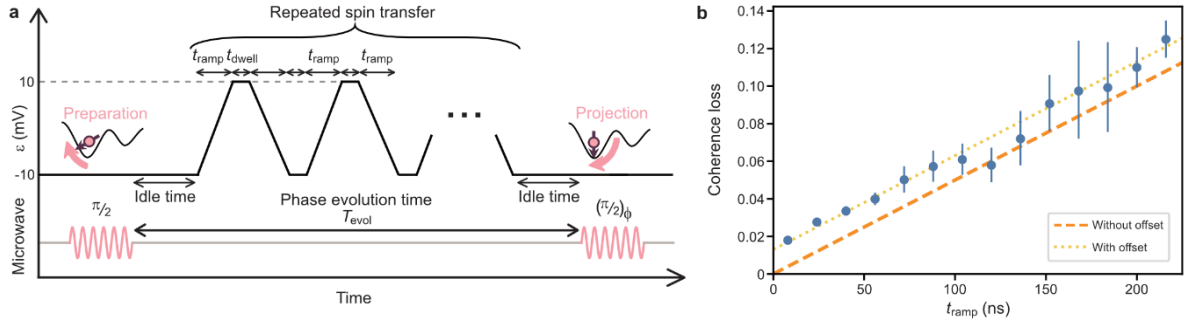

**Fig. S6. Ramp time dependence. (a)** Pulse schematic used for ramp time dependence measurement. We change the ramp time denoted by $t_{ramp}$, while fixing the dwell time to 48 ns. The idle time before and after the repeated transfer ramps is adjusted to keep $T_{evol}$ constant. **(b)** Coherence loss as a function of $t_{ramp}$. The orange dashed (yellow dotted) line shows a linear guide to the eye without (with) a 1.3 % offset.